\documentstyle[11pt]{article}

\makeatletter
\@addtoreset{equation}{section}
\makeatother

\def\dalemb#1#2{{\vbox{\hrule height .#2pt
        \hbox{\vrule width.#2pt height#1pt \kern#1pt
                \vrule width.#2pt}
        \hrule height.#2pt}}}

\def\td{\tilde}

\let\a=\alpha

 \def\bd{\begin{document}} \def\ed{\end{document}}
\def\ds{\documentstyle} \let\fr=\frac \let\bl=\bigl \let\br=\bigr
\let\Br=\Bigr \let\Bl=\Bigl
\let\bm=\bibitem
\let\na=\nabla
\let\pa=\partial \let\ov=\overline
\newcommand{\be}{\begin{equation}}
\newcommand{\ee}{\end{equation}}
\def\bog{Bogomol'nyi\ }
\def\ba{\begin{array}}
\def\ea{\end{array}}
\def\ft#1#2{{\textstyle{{\scriptstyle #1}\over {\scriptstyle #2}}}}
\def\fft#1#2{{#1 \over #2}}
\def\del{\partial}
\def\sst#1{{\scriptscriptstyle #1}}
\def\oneone{\rlap 1\mkern4mu{\rm l}}
\newcommand{\ho}[1]{$\, ^{#1}$}
\newcommand{\hoch}[1]{$\, ^{#1}$}
\newcommand{\bea}{\begin{eqnarray}}
\newcommand{\eea}{\end{eqnarray}}
\newcommand{\ra}{\rightarrow}
\newcommand{\lra}{\longrightarrow}
\newcommand{\Lra}{\Leftrightarrow}
\newcommand{\ap}{\alpha^\prime}
\newcommand{\bp}{\tilde \beta^\prime}
\newcommand{\tr}{{\rm tr} }
\newcommand{\Tr}{{\rm Tr} }
\newcommand{\NP}{Nucl. Phys. }
\newcommand{\tamphys}{\it Center for Theoretical Physics,
Texas A\&M University, College Station, Texas 77843}

\newcommand{\auth}{H. L\"u and C.N. Pope}

\begin{document}

\hfill{CTP-TAMU-03/96}

\hfill{hep-th/9601089}

\vspace{20pt}

\begin{center}
{ \large {\bf An approach to the classification of $p$-brane 
solitons}}

\vspace{30pt}
\auth

\vspace{15pt}
{\tamphys}
\vspace{30pt}

\underline{ABSTRACT}
\end{center}
\vspace{40pt}

We give a review of some recent work on the construction and
classification of $p$-brane solutions in maximal supergravity theories in
all dimensions $4\le D\le 11$.  These solutions include isotropic elementary
and solitonic $p$-branes, dyonic $p$-branes, and multi-scalar $p$-branes. 
These latter two categories include massless strings and black holes as
special cases.  For all the solutions, we analyse their residual unbroken
supersymmetry by means of an explicit construction of the eigenvalues of the
\bog matrix, defined as the anticommutator of the conserved supercharges.

{\vfill\leftline{}\vfill
\vskip 10pt
\footnoterule
{\footnotesize	Research supported in part by DOE
Grant DE-FG05-91-ER40633 \vskip	-12pt}}

\pagebreak
\setcounter{page}{1}

\section{Introduction}

     The study of $p$-brane solitons in low-energy superstring theories has
been the subject of detailed investigation recently [1--12].  With the
growing realisation that the duality symmetries of string theories can
bring about a unification of theories that were previously thought to
have been distinct, and that the non-perturbative spectrum of string
theory is within grasp, it becomes more and more important to achieve
a thorough understanding of the solitonic states, since they will have
an important r\^ole to play in the emerging theory.  Many solitonic
$p$-brane solutions have been found, and in this review we shall not
attempt to cover all of the ground.  Instead, we shall focus on an
approach to the problem of classifying the solutions that we have been
developing, which seems to have an advantage of simplicity.  We also
find a simple way of determining the supersymmetry of the solutions. 

     The class of low-energy effective actions that we shall discuss are
those obtained by dimensional reduction of the effective action of
the ten-dimensional type IIA string theory.  Equivalently, and more
simply, these maximal supergravity theories can be viewed as coming
from  eleven-dimensional supergravity.  As we shall see, all the
resulting lower-dimensional supergravity theories can be discussed in
a unified way, and the supersymmetry of the $p$-brane solutions in any
dimension can   easily be determined by simply using the properties
of the  eleven-dimensional transformation rules.  In particular, there
is no need to examine the supersymmetry transformation rules of the
lower-dimensional theories themselves, and there is no need to
decompose the  eleven-dimensional gamma matrices into products of
lower-dimensional ones.

     The bosonic sector of the Lagrangian for $D=11$ supergravity is 
${\cal L}= \hat e \hat R -\ft1{48} \hat e\, \hat F_4^2 +\ft16 
\hat F_4\wedge\hat F_4\wedge \hat A_3$, where $\hat A_3$ is the 3-form
potential for the 4-form fields strength $\hat F_4$.  Upon Kaluza-Klein
dimensional reduction to $D$ dimensions, this yields the following 
Lagrangian:
\bea
{\cal L} &=& eR -\ft12 e\, (\del\vec\phi)^2 -\ft1{48}e\, e^{\vec a\cdot 
\vec\phi}\, F_4^2 -\ft{1}{12} e\sum_i 
e^{\vec a_i\cdot \vec\phi}\, (F_3^{i})^2
-\ft14 e\, \sum_{i<j} e^{\vec a_{ij}\cdot \vec\phi}\, (F_2^{ij})^2
\nonumber\\
&& -\ft14e\, \sum_i e^{\vec b_i\cdot \vec\phi}\, ({\cal F}_2^i)^2
-\ft12 e\, \sum_{i<j<k} e^{\vec a_{ijk} \cdot\vec \phi}\,
(F_1^{ijk})^2 -\ft12e\, \sum_{i<j} e^{\vec b_{ij}\cdot \vec\phi}\,
({\cal F}_1^{ij})^2 + {\cal L}_{\sst{FFA}}\ ,\label{dgenlag}
\eea
where $F_4$, $F_3^i$, $F_2^{ij}$ and $F_1^{ijk}$ are the 4-form, 3-forms,
2-forms and 1-forms coming from the dimensional reduction of $\hat F_4$ in
$D=11$; ${\cal F}_2^i$ are the 2-forms coming from the dimensional reduction
of the vielbein, and ${\cal F}_1^{ij}$ are the 1-forms coming from the
dimensional reduction of these 2-forms.  The quantity $\vec\phi$ denotes an
$(11-D)$-component vector of scalar fields, which we shall loosely refer to
as dilatonic scalars that arise from the dimensional reduction of the
elfbein.  These scalars appear undifferentiated, via the exponential
prefactors for the antisymmetric-tensor kinetic terms, and they should be
distinguished from the remaining spin-0 quantities in $D$ dimensions, namely
the 0-form potentials $A_0^{ijk}$ and ${\cal A}_0^{ij}$.  These fields
have constant shift symmetries, under which the action is invariant, and are
thus properly thought of as 0-form potentials rather than true scalars.  In
particular, their 1-form field strengths can adopt topologically non-trivial
configurations, corresponding to magnetically-charged sources for $p$-brane
solitons.

     There are two subtleties that need to be addressed with regard
to the $p$-brane solutions following from (\ref{dgenlag}).  Firstly, the
dimensional reduction of the $\hat F_4\wedge\hat F_4\wedge \hat A_3$ term in
$D=11$ gives rise to a term in $D$ dimensions, which we denote by 
${\cal L}_{\sst{FFA}}$.  This gives contributions to the equations of motion
which, in general, will be non-zero.  The solutions that we wish to discuss, 
in common with most of those in the literature, are ones for which this term
can be ignored.  Thus we must check that our solutions are such that the
extra terms in the equations of motion following from 
${\cal L}_{\sst{FFA}}$
vanish.  The second complication is that the various field strengths occurring in (\ref{dgenlag})
are not, in general, simply given by the exterior derivatives of potentials.
There are extra terms, which we refer to loosely as ``Chern-Simons terms,''
coming from the process of dimensional reduction.  Again, the solutions of
interest will be ones where these extra terms vanish, and so again, this
means that certain constraints must be satisfied in order for this to be
true.  Both of these issues are discussed in detail in \cite{lp1} and so we
shall not consider them further here.

     The only aspect of the bosonic Lagrangians (\ref{dgenlag}) 
that remains unexplained is the constant vectors $\vec a_{i\cdots j}$ and
$\vec b_{i\cdots j}$ appearing in the exponential prefactors of the kinetic terms for the
antisymmetric tensors.  As was shown in \cite{lp1}, these ``dilatonic
vectors'' can be expressed  as follows:
\bea
&&F_{\sst{MNPQ}}\qquad\qquad\qquad\qquad\qquad\qquad
{\rm vielbein}\nonumber\\
{\rm 4-form:}&&\vec a = -\vec g\ ,\nonumber\\
{\rm 3-forms:}&&\vec a_i = \vec f_i -\vec g \ ,\label{dilatonvec}\\
{\rm 2-forms:}&& \vec a_{ij} = \vec f_i + \vec f_j - \vec g\ ,
\qquad\qquad\qquad \,\,\, \,\vec b_i = -\vec f_i\nonumber\,\\
{\rm 1-forms:}&&\vec a_{ijk} = \vec f_i + \vec f_j + \vec f_k -\vec g
\ ,\qquad\qquad\vec b_{ij} = -\vec f_i + \vec f_j\ ,\nonumber
\eea
where the vectors $\vec g$ and $\vec f_i$ have $(11-D)$ components
in $D$ dimensions, and are given by
\bea
\vec g &=&3 (s_1, s_2, \ldots, s_{11-\sst D})\ ,\nonumber\\
\vec f_i &=& \Big(\underbrace{0,0,\ldots, 0}_{i-1}, (10-i) s_i, s_{i+1},
s_{i+2}, \ldots, s_{11-\sst D}\Big)\ ,
\eea
where $s_i = \sqrt{2/((10-i)(9-i))}$.  It is easy to see that they satisfy
\be
\vec g \cdot \vec g = \ft{2(11-D)}{D-2}, \qquad
\vec g \cdot \vec f_i = \ft{6}{D-2}\ ,\qquad
\vec f_i \cdot \vec f_j = 2\delta_{ij} + \ft2{D-2}\ .\label{gfdot}
\ee
Note that the definitions in (\ref{dilatonvec}) are given for $i<j<k$, and
that the vectors $\vec a_{ij}$ and $\vec a_{ijk}$ are antisymmetric in their
indices. The 1-forms ${\cal F}_{\sst{M}i}^{(j)}$ and hence the vectors
$b_{ij}$ are only defined for $i<j$, but it is sometimes convenient to
regard them as being antisymmetric too, by defining $\vec b_{ij}=-\vec b_{ji}$
for $i>j$.  Eqn.\ (\ref{dilatonvec}), together
with (\ref{gfdot}), contains all the necessary information about the dilaton
vectors in $D$-dimensional maximal supergravity.

     In the absence of a $p$-brane, the equations of motion following from 
(\ref{dgenlag}) admit a $D$-dimensional Minkowski vacuum, with $SO(1,D-1)$
Lorentz symmetry.  The $p$-brane solutions we shall consider break this to
$SO(1,d-1)\times SO(D-d)$, where $d=p+1$ is the dimension of the world     
volume of the $p$-brane.  The metric tensor, compatible with this residual
symmetry, is given by 
\be
ds^2 = e^{2A}\, dx^\mu dx^\nu \eta_{\mu\nu} + e^{2B}\, dy^m dy^m\ ,
\ee
where $A$ and $B$ are functions only of $r=\sqrt{y^m y^m}$.  The coordinates
$x^\mu$ lie in the $d$-dimensional world volume, and $y^m$ lie in the
$(D-d)$-dimensional transverse space.  

     The solutions that we shall consider involve a subset of the
dilatonic scalars $\vec\phi$, and a subset of the antisymmetric tensor field
strengths $F$ and ${\cal F}$.  The rank $n$ of all the participating fields
strengths in a given solution are all the same, although in dimensions 
$D\le 7$ certain field strengths of higher rank will be dualised to field
strengths of lower rank.  Each participating field strength can have
non-vanishing components given by either
\be
F_{m\mu_1\ldots\mu_{n-1}} = \epsilon_{\mu_1\ldots\mu_{n-1}} (e^C)'\,
{y^m\over r}\qquad
{\rm or}\qquad
F_{m_1\ldots m_n} = \lambda \epsilon_{m_1\ldots m_np}\,
\fft{y^p}{r^{n+1}}\ ,\label{fansatz}
\ee
where a prime denotes a derivative with respect to $r$.  The first case
gives rise to an elementary $(d-1)$-brane with $d=n-1$, and the second
gives rise to a solitonic $(d-1)$-brane with $d= D-n-1$.  These two field
configurations have electric charge
$u=\ft{1}{4\omega_{\sst{D}-n}} \int_{\del\Sigma} *F$ or
magnetic charge $v=\ft{1}{4\omega_{n}}\int_{\del\Sigma} F$ respectively,
where $\del\Sigma$ is the boundary $(D-d-1)$-sphere of the transverse 
space.  The first case corresponds to an elementary, or fundamental,
$p$-brane solution, and the second case corresponds to a solitonic $p$-brane
solution.  In a case where the degree $n$ of the field strength is such that
$D=2n$, the possibility arises that field strengths might have both electric
and magnetic contributions, involving both of the ans\"atze in
(\ref{fansatz}) simultaneously.  In such cases, the configuration describes
a dyonic $p$-brane.  It is useful to distinguish two different kinds of
dyonic configuration.  In a dyonic solution of the first type, each
participating field strength has only electric or magnetic non-vanishing
components, but some of the field strengths are electric while others are
magnetic.  In a dyonic solution of the second type, each participating field
strength has both electric and magnetic non-vanishing components.  We shall
encounter both kinds of dyonic solution later.

     In order to determine the supersymmetry properties of the various
$p$-brane solutions, it suffices to study the transformation laws of $D=11$
supergravity.  In particular, from the commutator of the conserved
supercharges  $Q_\epsilon = \int_{\del\Sigma} \bar\epsilon \Gamma^{\sst{ABC}}
\psi_{\sst C}\, d\Sigma_{\sst{AB}}$, we may read off the $32\times32$ 
\bog matrix ${\cal M}$, defined by $[Q_{\epsilon_1},Q_{\epsilon_2}] =
\epsilon_1^\dagger {\cal M} \epsilon_2$, whose zero eigenvalues correspond to
unbroken components of $D=11$ supersymmetry. The expression for ${\cal M}$
for maximal supergravity in an arbitrary dimension $D$ then follows by
dimensional reduction of the expression in $D=11$.  A
straightforward calculation shows that it is given by \cite{lp1}
\bea
{\cal M} &=& m\oneone + u\, \Gamma_{012} + u_i\, \Gamma_{01i} +
\ft12 u_{ij}\, \Gamma_{0ij} +p_i \Gamma_{0i}\nonumber\\
&&+ v\,\Gamma_{\hat1\hat2\hat3\hat4\hat5} + v_i \,
\Gamma_{\hat1\hat2\hat3\hat4i}+\ft12 v_{ij}\, \Gamma_{\hat1\hat2\hat3ij}
+ \ft16 v_{ijk}\, \Gamma_{\hat1\hat2ijk} + q_i\, \Gamma_{\hat1\hat2\hat3 i}
+\ft12 q_{ij}\, \Gamma_{\hat1\hat2ij}\ .\label{genbog}
\eea
The indices $0, 1,\ldots$ run over the dimension of
the $p$-brane worldvolume, $\hat1,\hat2,\ldots$ run over the transverse
space of the $y^m$ coordinates, and $i,j,\ldots$ run over the dimensions
that were compactified in the Kaluza-Klein reduction from 11 to $D$
dimensions. The quantities $u$, $u_i$, $u_{ij}$ and $p_i$ are the 
electric Page charges associated with the field strengths $F_4$, $F_3^i$,
$F_2^{ij}$ and ${\cal F}_2^i$ respectively, while $v$, $v_i$, $v_{ij}$,
$v_{ijk}$, $q_i$ and $q_{ij}$ are the magnetic Page charges associated with 
$F_4$, $F_3^i$, $F_2^{ij}$, $F_1^{ijk}$, ${\cal F}_2^i$ and ${\cal F}_1^{ij}$
respectively.  The quantity $m$ is the mass per unit $p$-volume for the
solution, given by $\ft12 (A'-B')e^{-B} r^{\td d +1}$ in the limit
$r\rightarrow \infty$, where $\td d\equiv D-d-2$.

     Once the mass per unit $p$-volume and the Page charges have been
determined for a given $p$-brane solution, it becomes a straightforward
algebraic exercise to calculate the eigenvalues of the \bog matrix, given by 
(\ref{genbog}).  The fraction of $D=11$ supersymmetry that is preserved by
the solution is then equal to $k/32$, where $k$ is the number of zero
eigenvalues of the $32\times32$ \bog matrix. 

\section{Single-scalar solutions}

     In the single-scalar $p$-brane solutions, the Lagrangian (\ref{dgenlag})
is consistently truncated to the simple form 
\be
{\cal L} = e
R -\ft12 e\, (\del\phi)^2 -\fft1{2n!} e\, e^{a\phi} F_n^2 \ ,\label{genlag}
\ee
where the scalar field $\phi$ is some linear combination of the dilatonic
scalars $\vec\phi$ of the $D$-dimensional theory, and $F_n$ is a single
canonically-normalised $n$-index field strength, to which all of the $N$
original field strengths that participate in the solution are proportional.
The constant $a$ appearing in the exponential prefactor can conveniently be
parameterised as
\be
a^2 = \Delta - \fft{2d\tilde d} {D-2}\ ,\label{avalue}
\ee
where $\tilde d\equiv D-d -2$ and $d\tilde d = (n-1)(D-n-1)$.  The
quantity $\Delta$, unlike $a$ itself, is preserved under Kaluza-Klein
dimensional reduction.  

     If we write $\vec\phi=\vec n\, \phi + \vec\phi_\perp$, where $\vec n$ is
a constant unit vector and $\vec n\cdot \vec\phi_\perp=0$, then it is easy to
see that the conditions that must be satisfied in order for the truncation to
(\ref{genlag}) to be consistent with the equations of motion for
$\vec\phi_\perp$ are that $\vec a_\a\cdot \vec n =a$ for $\a$, where $\vec
a_\a$ ($1\le\a\le N$) denotes the dilaton vectors associated with the $N$
participating field strengths, and in addition, that
\be
\sum_{\alpha} \vec a_{\alpha} F_{\alpha}^2 = a \vec n \sum_\alpha
F^2_\alpha \ , {\rm \  and\ hence\ }
\sum_\alpha M_{\alpha\beta} F_\alpha^2 = a^2 \sum_\alpha F_\alpha^2\ ,
\label{fequation1}
\ee
where $M_{\alpha\beta}\equiv \vec a_\alpha \cdot \vec a_\beta$.  If the
matrix $M_{\alpha\beta}$ is invertible, we therefore have
\be
F_\beta^2 = a^2 \sum_\alpha (M^{-1})_{\alpha\beta} \sum_\gamma F_\gamma^2
\ ,\qquad 
a^2 = \Big( \sum_{\alpha, \beta} (M^{-1})_{\alpha\beta} \Big)^{-1}
\ ,\label{avaluesol}
\ee
which gives rise to a $p$-brane solution with $\Delta = \Big( \sum_{\alpha,
\beta} (M^{-1})_{\alpha\beta} \Big)^{-1} + 2d\tilde d/(D-2)$.  In this
case, we can easily see that $a \ne 0$, and hence the unit vector $\vec n$
can be read off from eqn (\ref{fequation1}).  If $M_{\alpha\beta}$ is
non-invertible, then it is clear from (\ref{fequation1}) that one solution is
is when $F_\a^2$ is a zero eigenvector of $M_{\alpha\beta}$, with the
constant $a$ being zero.  It turns out that this is the only solution in the
singular case that does not simply reduce to an already-considered 
non-singular case with a smaller
number $N$ of participating field strengths \cite{lp1}.  It is also clear
that if the number of participating field strengths exceeds the dimension
$(11-D)$ of the dilaton vectors, then the associated matrix $M_{\a\beta}$
will be singular, and in fact it turns out that in all such cases, there is
no new solution \cite{lp1}.  Thus in any dimension $D$, it follows that the
number $N$ of participating field strengths must always satisfy $N\le 
11-D$.

     Having reduced the Lagrangian (\ref{dgenlag}) to (\ref{genlag}) by the
above procedure, it is now a simple matter to obtain solutions for the
equations of motion that follow from (\ref{genlag}).  An important point is 
that one can reduce the second-order equations of motion for $A, B$ and 
$\phi$ to first-order equations, by making an ansatz where they are all 
proportional, and in which the 
exponential factor in these variables, coming from the $F^2$ source term on 
the right-hand sides of their equations of motion, is required to be 
proportional to $A'$ \cite{lpss}.  In the case of solutions where some 
supersymmetry remains unbroken, these requirements are dictated by the  
supersymmetry transformation rules.  It also provides a way to obtain 
solutions of the same general form, even in cases where there is no 
residual unbroken supersymmetry.  The solution is
\bea
ds^2 &=& \Big (1+\fft{k}{r^{\tilde d}}\Big)^{-\ft{4\tilde d}{\Delta(D-2)}}
\, dx^\mu dx^\nu \eta_{\mu\nu} + \Big(1+ \fft{k}{r^{\tilde d}}\Big)^{
\ft{4d}{\Delta(D-2)}}\, dy^m dy^m\ ,\nonumber\\
e^\phi &=& \Big (1+\fft{k}{r^{\tilde d}}\Big)^{
\ft{2a}{\epsilon \Delta}}\ , \label{gensol}
\eea
where $\epsilon = 1$ and $-1$ for the elementary and the
solitonic solutions respectively, and $k= - \sqrt{\Delta}
\lambda/(2 \tilde d)$.  In the elementary case, the function $C$ satisfies
the equation
\be
e^C = \fft{2}{\sqrt\Delta} \Big ( 1+ \fft{k}{r^{\tilde d}}\Big)^{-1}\ .
\ee
The mass of the solution is given by $\lambda/(2\sqrt\Delta)$.  
Note that the dual of the solution for the field strength in the elementary
case is identical to the field strength of the solitonic case, and {\it vice
versa}.  For this reason, we shall only consider solutions for field
strengths with $n\le D/2$.

     In order to enumerate all the single-scalar $p$-brane solutions of this
type, one simply has to consider, for each dimension $D$ and each degree $n$
for the field strengths, all possible choices of the associated $N\le 11-D$ 
dilaton vectors, and then calculate the values of $a$, and the corresponding
ratios of participating field strengths, using the above equations.  This is
easily done for $n=4$ (where there is always only one field strength) and
$n=3$ (where the number of field strengths is small).  For $n=2$ and $n=1$,
where the numbers of field strengths grow significantly with decreasing
dimension $D$, the enumeration is most conveniently carried out by computer.
Substituting the results for the field strengths into (\ref{genbog}), it is
then straightforward, for each solution, to determine the fraction of the
original supersymmetry that is preserved.  Most of the solutions turn out to
break all of the supersymmetry, but in certain cases, some of the
supersymmetry survives.  Details may be found in \cite{lp1}; here, we
summarise the results:

\bigskip
\noindent{\it 4-Form solutions}
\medskip

   Since there is only one 4-form field strength in any dimension, there is a
a unique solution for each dimension $D$, in which the scalar field $\phi$
appearing in (\ref{genlag}) is taken to be the canonically-normalised scalar
proportional to the entire exponent of the prefactor for the 4-form's kinetic
term.  The value of $a$ for this solution corresponds to $\Delta=4$ for all
$D$.  The eigenvalues of the \bog matrix (\ref{genbog}) turn out to be 
$\mu=m \{0_{16}, 2_{16}\}$, where the subscripts denote the degeneracies 
of each eigenvalue. Thus these solutions preserve $\ft12$ of the $D=11$ 
supersymmetry.

\bigskip
\noindent{\it 3-Form solutions}
\medskip

     From (\ref{dilatonvec}) and (\ref{avaluesol}), it is easy to see that 
there exist solutions involving $N$ participating field strengths, with
values of $\Delta$ given by $\Delta=2+2/N$, where, as always, $N\le 11-D$.
Substituting the solutions for the ratios of the field strengths, and hence
the ratios of the Page charges, into (\ref{genbog}), one finds \cite{lp1}
that all the supersymmetry is broken except when $N=1$, and hence
$\Delta=4$, and the eigenvalues are the same as for the case of 4-form
solutions, namely $\mu=m \{0_{16}, 2_{16}\}$.  In this supersymmetric case, 
any one of the 3-form field strengths can be used in constructing the 
solution.  In $D=6$, there are further solutions,  associated with the 
possibility of truncating the theory to self-dual supergravity.  We shall 
discuss these further in the next section.

\bigskip
\noindent{\it 2-Form solutions}
\medskip

     As one can see from (\ref{dilatonvec}), the set of possible choices for
dilaton vectors is considerably larger for 2-form field strengths, and
correspondingly there are many more solutions.  If only a single 2-form field
strength participates in the solution, then as usual we find $\Delta=4$.
For $N=2$, we can have $\Delta=3$ or $\Delta=2$.  For $N=3$, there are always
solutions with $\Delta=\ft83$ and $\ft{12}7$.  In addition, if $D\le 5$,
there is a further solution with $\Delta=\ft43$.  As $N$ increases to its
maximum allowed value of $N=7$ in $D=4$, more and more solutions appear. 
Full details may be found in \cite{lp1}.  Here, we shall just give details
for the supersymmetric solutions.  These occur for $N=1,2,3$ and 4
participating field strengths, with
$\Delta=4/N=4,2,\ft43$ and 1 respectively.  The eigenvalues of the \bog
matrix are as follows:
\bea
\Delta=4: &&\mu=2 m \{0_{16},1_{16}\} \ ,\qquad\qquad D\le10\ ,\nonumber\\
\Delta=2: &&\mu=m \{0_{8},1_{16}, 2_{8}\} \ ,\qquad\qquad D\le9\ ,
\nonumber\\
\Delta=\ft43: &&\mu=\ft23 m \{0_{4},1_{12},2_{12},3_4\}\ ,
\qquad\qquad D\le5\ ,\label{n2eig}\\
\Delta=1': &&\mu=m \{0_{4},1_{24}, 2_{4}\}\ ,\qquad\qquad D=4 \ ,\nonumber
\eea
where we also indicate the maximum dimension $D$ in which each solution can
occur. We see that the $p$-brane solutions preserve $\ft12, \ft14,
\ft18$, and
$\ft18$ of the $D=11$ supersymmetry respectively.  The prime on the value of
$\Delta$ for the fourth case is for later convenience, to distinguish it from
another $\Delta=1$ configuration that will arise in the discussion of 1-form
solutions.  It should also be noted that the equations (\ref{avaluesol})
only determine the relations between the {\it squares} of the Page charges,
and thus the signs can be arbitrarily chosen.  For the configurations with
$N\le3$ participating field strengths that give rise to supersymmetric
solutions, the eigenvalues of the \bog matrix are insensitive to these sign
choices.  For the $N=4$ case, on the other hand, the choice of relative
signs does matter, and it turns out that there are exactly two possible sets
of eigenvalues that can result.  One of these is the supersymmetric solution
with $\Delta=1'$ listed in (\ref{n2eig}); for the inequivalent choice of
relative signs, the eigenvalues are $\mu=m \{ 1_{16},3_{16}\}$, and thus
there is no supersymmetry. Note that any one of the 2-form field strengths
can be used to obtain the $\Delta=4$ solution. For $\Delta=2$, there are
various possible pairs of field strengths that can be chosen, the number of
such choices increasing with decreasing $D$. An example, which can be used
in all dimensions $D\le9$, is to choose $F_2^{12}$ and ${\cal F}_2^{1}$,
with Page charges given by $u_{12}=p_1=\lambda/(4\sqrt2)$.  The
non-supersymmetric $\Delta=3$ solution  is obtained for other choices of two
participating field strengths, for  example $p_1=p_2=\lambda/(4\sqrt2)$. 
Examples with $N=3$ and $N=4$ field  strengths, yielding the $\Delta=\ft43$
and $\Delta=1$ supersymmetric  solutions, are
$u_{12}=u_{34}=u_{56}=\lambda/8$, and $u_{23}=u_{46}=u_{57}=p_1^*=\lambda/8$
respectively.  In the latter case, $p_1^*$ denotes an electric Page charge
for the dual of the field strength ${\cal F}_2^{1}$.  As mentioned above, if
the signs of the charges in this $\Delta=1$ case are changed, then eight of
the sign choices give the same solution whilst for the other eight we still
get a solution, with the same metric, but with  no supersymmetry preserved. 

\bigskip
\noindent{\it 1-Form solutions}
\medskip

     The situation for 1-form solutions is more complicated again; further
details may be found in \cite{lp1}.  As $N$ increases up to 7, there is a
considerable proliferation of solutions, mostly non-supersymmetric.  Here, we
shall just describe the supersymmetric $p$-brane solitons.  There are in
total eight inequivalent field configurations that can give rise to
supersymmetric solutions, namely one for each value of $N$ in the range $1\le
N\le 7$, together with a second inequivalent possibility for $N=4$.  In all
cases, the value of $\Delta$ is given by $\Delta=4/N$.  Four of the eight
solutions have \bog matrices with identical eigenvalues to those given in
(\ref{n2eig}) for 2-form solutions (although the \bog matrices themselves are
of course different).  These again arise for $N=1, 2, 3$ and 4 participating 
field strengths.  Examples of the Page charges that can give rise to these 
four case are $q_{12}$, $\{q_{12}, v_{123}\}$, $\{q_{12}, q_{45}, v_{123} 
\}$ and $\{ q_{12}, q_{45}, v_{123}, v_{345}\}$ respectively, where in each 
case all the listed charges are equal.  The remaining four solutions, which
cannot occur above $D=4$ dimensions, have eigenvalues as follows: 
\bea 
\Delta =1:&& \{q_{12}, q_{34}, q_{56}, v_{127}\}\ ,\qquad 
\mu =\ft12 m\{0_2, 1_8, 2_{12}, 3_8, 4_2\}\ ,\nonumber\\
\Delta = \ft45: && \{q_{12}, q_{34}, q_{56}, v_{127}, v_{347}\}\ ,\qquad
\mu= \ft25 m\{ 0_2, 1_2, 2_{12}, 3_{12}, 
4_2, 5_{2}\}\ ,\nonumber\\
\Delta = \ft23: && 
\{q_{12}, q_{34}, q_{56}, v_{127}, v_{347}, v_{567}\}\ ,\qquad
\mu=\ft13m \{ 0_2, 2_6, 3_{16}, 4_6, 
6_2\}\ ,\label{n1eig}\\
\Delta =\ft47:&& 
\{q_{12}, q_{34}, q_{56}, v_{127}, v_{347},v_{567}, v_7^*\}\ ,\qquad
\mu=\ft27 m \{0_2, 3_{14}, 4_{14}, 7_2\}\ ,\nonumber
\eea
where in each case the listed set of Pages charges provides an example of a 
set that gives rise to the solution.  In each case, the Page charges are all 
equal.  All of these four cases correspond to solutions that preserve
$\ft1{16}$ of the $D=11$ supersymmetry.  In the first of these four
solutions, the eigenvalues are insensitive to the relative sign choices for
the Page charges, but in the last three cases, we again find the phenomenon
that there are precisely two inequivalent sets of eigenvalues for each
$\Delta$, depending on the relative signs of the Page charges.  We have
given the choices that include zero eigenvalues.  The other choices, for
which there is no supersymmetry, give rise to the sets of eigenvalues 
\bea 
\Delta = \ft45: && 
\mu= \ft25 m\{ 1_8, 2_8, 3_8, 4_8 \}\ ,\nonumber\\
\Delta = \ft23: && 
\mu=\ft13m \{ 1_4, 2_8, 3_8, 4_8, 
5_4\}\ ,\label{n1eigns}\\
\Delta =\ft47:&& 
\mu=\ft27 m \{1_2, 2_6, 3_8, 4_8,5_6,6_2\}\ ,\nonumber
\eea

     To conclude this section, we give a table that summarises the dimensions
in which the various supersymmetric $p$-brane solutions first occur:
\bigskip\bigskip

\centerline{
\begin{tabular}{|c|c|c|c|c|}\hline
&\phantom{for} 4-form\phantom{for} &\phantom{for} 3-form\phantom{for}
&\phantom{for} 2-form\phantom{for} & 1-form\\ \hline\hline
$D=11$  & $\Delta = 4 $ &        &        &       \\  \hline
$D=10$   &               & $\Delta =4$ & $\Delta=4$  &   \\ \hline
$D=9$    &    &     & $\Delta =2$ & $\Delta =4$ \\ \hline
$D=8$    &    &     &  & $\Delta =2$ \\  \hline
$D=7$    &    &     &  & \\  \hline
$D=6$    &    &  $\Delta =2$   &  & $\Delta = \ft43,1'$ \\  \hline
$D=5$    &    &     & $\Delta=\ft43$ & \\  \hline
$D=4$    &    &  &$\Delta = 1'$ & $\Delta = 1, \ft45, \ft23, \ft47$\\
\hline
\end{tabular}}

\bigskip
\centerline{Table 1: Supersymmetric $p$-brane solutions}

\bigskip

\section{Dyonic solutions}

     In dimensions $D=2n$, the field strength $F_n$ can in principle have
components given by both the elementary and solitonic ans\"atze 
(\ref{fansatz}) simultaneously.  In this case, the equations of motion 
can be reduced to the two independent differential equations
\be
\phi'' + n \fft{\phi'}{r} =\ft12 a (S_1^2 - S_2^2)\ ,\qquad
A'' + n \fft{A'}{r} = \ft14 (S_1^2 + S_2^2)\ ,\label{phiadiff}
\ee
together with the relations $B=-A$, $(e^C)' = \lambda_1 e^{a\phi + 2(n-1)A}
r^{-n}$, where $S_1$ and $S_2$ are given by
\be
S_1 = \lambda_1 e^{\ft12 a \phi + (n-1)A} r^{-n}\ ,\qquad
S_2 = \lambda_2 e^{-\ft12 a\phi + (n-1)A} r^{-n}\ .\label{s1s2}
\ee
By making a suitable ansatz of the kind discussed in the previous section, 
which reduces the equations to first-order equations, one finds \cite{lp1}
that the equations (\ref{phiadiff}) admit a simple solution either when
$a^2=n-1$, and hence $\Delta=2n-2$, given by
\be
e^{-\ft12 a \phi -(n-1) A} = 1 + \ft{\lambda_1}{a\sqrt2} r^{-n+1}\ ,\qquad
e^{+\ft12 a \phi -(n-1) A} = 1 + \ft{\lambda_2}{a\sqrt2} r^{-n+1}\ ,
\label{dyonicsol1}
\ee
or when $a=0$, and hence $\Delta=n-1$, given by
\be
\phi = 0\ ,\qquad e^{- (n-1) A} = 1 +\ft12
\sqrt{\ft{\lambda_1^2+\lambda_2^2}{(n-1)}} r^{-n+1}\ .\label{dyonicsol2}
\ee

     The possible dimensions for dyonic solutions are $D=8$, 6 and 4.  As
discussed in \cite{lp1}, solutions of the kind we are discussing, where the
contributions from the ${\cal L}_{\sst{FFA}}$ term are assumed to vanish,
cannot occur in $D=8$.  Furthermore, the solutions (\ref{dyonicsol1}) and 
(\ref{dyonicsol2}) would require that the 4-form have a dilaton prefactor
with $\Delta=6$ or $\Delta=3$, whereas in fact it has $\Delta=4$.  Thus we
are left with the cases $D=6$ and $D=4$.

\bigskip
\noindent{\it Dyonic solutions in  $D=6$}
\medskip

     In $D=6$, the constraints implied by the requirement that ${\cal
L}_{\sst{FFA}}$ not contribute imply that there cannot be dyonic solutions of
the first type, where each individual field strength has only electric or
magnetic components.  However, dyonic solutions of the second type, where
each field strength has electric and magnetic components, can occur, for
$N\le5$ participating field strengths, with $\Delta=2+2/N$.  The simple
dyonic solutions (\ref{dyonicsol1}) and (\ref{dyonicsol2}) require $\Delta=4$
and $\Delta=2$ respectively, and thus we see that there exist dyonic
solutions given by (\ref{dyonicsol1}) when $N=1$.  
The mass per unit length of this
dyonic string is $m=\ft14(\lambda_1 +\lambda_2)$, and the Page charges are
$u=\ft14\lambda_1$ and $v=\ft14 \lambda_2$.  The eigenvalues of the
Bogomol'nyi matrix are given by
\be
\mu = m\pm u \pm v = \{0_8, (\ft12 \lambda_1)_8,
(\ft12 \lambda_2)_8, (\ft12 (\lambda_1 + \lambda_2))_8 \}\ ,\label{extra1}
\ee
where, as usual, the subscripts on the eigenvalues indicate their
degeneracies.  Thus the solution preserves $\ft14$ of the supersymmetry
\cite{dfkr}.  When either $\lambda_1=0$ or $\lambda_2=0$, the solution
reduces to the previously-discussed purely solitonic and purely
elementary solutions, which preserve $\ft12$ of the supersymmetry.  When
$\lambda_1=\lambda_2$, in which case the field strength becomes self-dual
and the dilaton vanishes, the solution is equivalent to
the self-dual string in $D=6$ self-dual supergravity, which we shall discuss
below.  When $\lambda_1 =-\lambda_2$, the field strength is anti-self-dual,
and we have a massless string which preserves $\ft12$ of the supersymmetry;
however, the eigenvalues, given by (\ref{extra1}), for such a solution are
not positive semi-definite.   In this case, the dilaton field does not
vanish, and hence the solution is distinct from the anti-self-dual string in
$D=6$ anti-self-dual gravity.  It is worth remarking that the eigenvalues
(\ref{extra1}) for these dyonic solutions of the second type are quite
different from those for all the solutions we have discussed previously.
In those cases, the eigenvalues are non-negative as long as the mass per
unit $p$-volume is positive.  However, for the dyonic solutions of the
second type, we see that the Page charges can be chosen so that eigenvalues
(\ref{extra1}) of the Bogomol'nyi matrix take both signs, even when the
mass is positive.

     In the above discussion, we saw that the field strength of the solution
could be chosen to be either self-dual or anti-self-dual.  In fact, one can
alternatively truncate the theory so as to retain a single 3-form
field strength on which a self-dual (or anti-self-dual) condition is imposed
\cite{romans}. In this case, the dilatonic fields are all consistently
truncated from the theory, implying that $a=0$ and hence $\Delta =2$.  The
metric is given by (\ref{dyonicsol2}) with
$\lambda_1=\lambda_2=\lambda$. The mass per unit length is given by $m=\ft12
\sqrt{(\lambda_1^2 +
\lambda_2^2)/\Delta} = \ft12 \lambda$; the Page charges of the solution
comprise an electric charge $u$ and a magnetic charge $v$, with $u=v=\ft14
\lambda$ ($v=-\ft14 \lambda$ for the anti-self-dual case). The
eigenvalues of the Bogomol'nyi matrix are given by $\mu = \fft{\lambda}{4}
(2\pm 1\pm1)= m \{0_8,1_{16}, 2_8\}$, and so a quarter of the $D=11$
supersymmetry is preserved in this (anti)-self-dual case.  Note that the
mass per unit $p$-volume of the self-dual (anti-self-dual) solution in the
previous paragraph is given by $m=u + v$, whilst the mass of these
solution in $D=6$ self-dual (anti-self-dual) supergravity is given by
$m=\sqrt{u^2 + v^2}$.

\bigskip
\noindent{\it Dyonic solutions in  $D=4$}
\medskip

     In $D=4$, the solutions (\ref{dyonicsol1}) and (\ref{dyonicsol2}) occur
when $\Delta=2$ and $\Delta=1$, corresponding to $a=1/\sqrt3$ and $a=0$.  
As we discussed for the elementary and solitonic solutions in the previous
section, in the case $\Delta=1$, arising when there are 4 participating 
field strengths, the relative signs of the Page charges, which are 
undetermined by the field equations, affect the eigenvalues of the \bog 
matrix, giving two inequivalent outcomes, one with supersymmetry and the 
other without.  We can therefore now have three inequivalent sets of
eigenvalues, depending on whether the relative signs in each of the electric
and the magnetic sectors are chosen both to reduce to the previous
supersymmetric choices, or else one supersymmetric and the other not, or
finally both non-supersymmetric. (The occurrence of non-supersymmetric
solutions for certain sign choices for the Page charges has also been seen
recently in \cite{ko}.) Accordingly, we list the three possibilities for
$\Delta=1$ below, in this order: 
\bea
\Delta=1: && m=\ft12   \lambda\ ,\qquad
\{u_{57}, u_{46}, u_{23}, p_1^*, v_{57}, v_{46}, v_{23}, q_1^*\}=
\ft18 \{\lambda_1, \lambda_1, \lambda_1, \lambda_1,
\lambda_2, \lambda_2, \lambda_2, -\lambda_2\}\ ,\nonumber\\
&&\mu=\ft12 \lambda\, \{0_4, 1_{24}, 2_4\}\ ,
\nonumber\\
&&\nonumber\\
\Delta=1: && m=\ft12   \lambda\ ,\qquad
\{u_{57}, u_{46}, u_{23}, p_1^*, v_{57}, v_{46}, v_{23}, q_1^*\}=
\ft12 \{\lambda_1, \lambda_1, \lambda_1, \lambda_1,
\lambda_2, \lambda_2, \lambda_2, \lambda_2\}\ ,\nonumber\\
&&\mu=\ft12  \{(\lambda-\ft12 \lambda_2)_4,
 (\lambda +\ft12 \lambda_2)_4,
 (\lambda -\sqrt{\lambda_1^2+\ft14\lambda_2^2})_{12},
 (\lambda +\sqrt{\lambda_1^2+\ft14\lambda_2^2})_{12} \}\ ,
\nonumber\\
&&\nonumber\\
\Delta=1: && m=\ft12   \lambda\ ,\qquad
\{u_{57}, u_{46}, u_{23}, p_1^*, v_{57}, v_{46}, v_{23}, q_1^*\}=
\ft12 \{\lambda_1, \lambda_1, \lambda_1, -\lambda_1,
\lambda_2, \lambda_2, \lambda_2, \lambda_2\}\ ,\nonumber\\
&&\mu=\ft14 \lambda\, \{1_{16}, 3_{16} \}\ ,
\nonumber\\
&&\nonumber\\
\Delta = 2:&&m=\fft{\lambda_1 + \lambda_2}{2\sqrt2}\ ,\qquad
\{p_1, u_{12}, q_1, v_{12} \} = \ft{1}{4\sqrt{2}}
\{\lambda_1,\lambda_1, \lambda_2, \lambda_2\}\ ,\label{dyonicd4}\\
&&\mu = \sqrt2\{(\lambda_1 + \lambda_2 - \lambda)_8,\,
(\lambda_1 + \lambda_2 + \lambda)_8,\,
(\lambda_1 + \lambda_2)_{16}\}\ ,\nonumber
\eea
where $\lambda\equiv \sqrt{\lambda_1^2 + \lambda_2^2}$.
The first $\Delta=1$ solution always preserves $\ft18$ of the
supersymmetry, regardless of the values of $\lambda_1$ and $\lambda_2$. 
The second $\Delta=1$ solution breaks all the supersymmetry for generic 
values of the Page charges, but gives rise to a supersymmetric elementary 
solution when $\lambda_2=0$.  The last $\Delta=1$ solution breaks all the 
supersymmetry for all values of the Page charges. For $\Delta=2$, we can
have zero eigenvalues only for the following three cases: $\lambda_1=0$,
$\lambda_2=0$ or $\lambda_1 = -\lambda_2$.  The first two cases correspond
to the purely solitonic and purely elementary solutions which preserve
$\ft14$ of the supersymmetry. The third case gives rise to a massless black
hole (which has been discussed in \cite{cvetic}), which preserves $\ft12$ of
the supersymmetry.  However, some of the eigenvalues are negative in this
case.

\section{Multi-scalar solutions}

     The third class of solutions that we shall describe here are ones where
the individual Page charges of the field strengths participating in a
solution are independent parameters (unlike the solutions discussed in the 
previous two sections, where they are fixed in a specific set
of ratios in any given solution).  This is
achieved by allowing more than one independent combination of the $(11-D)$
dilatonic scalar fields to be excited in the solution.  In fact, the number
of independent combinations is precisely equal to the number $N$ of
independent Page charges.  These combinations can be expressed as
$\varphi_\a=\vec a_\a\cdot \vec\phi$, where as usual $\vec a_\a$ denotes the
set of $N$ dilaton vectors for the $N$ participating field strengths.  One
can easily verify from the equations of motion following from
(\ref{dgenlag}) that the remaining orthogonal combinations of the $\vec\phi$
fields can be consistently set to zero.  A natural choice for the solutions
turns out to be to take $d A +\td d B=0$ and $A=- {\epsilon \td d \over D-2}
\sum_{\a,\beta} (M^{-1})_{\a\beta}\, \varphi_\beta$. The remaining equations
can then be solved by making an ansatz that is analogous to the one 
discussed in section 2, which reduces the second-order equations of motion 
to first-order equations \cite{lp2}.  For this ansatz to be consistent, it 
turns out that the matrix $M_{\a\beta}$ of dot products of dilaton vectors 
must take the special form $M_{\a\beta}=4\delta_{\a\beta} -2 d \td d/(D-2)$. 
In fact this is precisely the same as the form that $M_{\a\beta}$ takes in 
the various supersymmetric single-scalar solutions that we discussed in 
section 2.  Thus we conclude that multi-scalar solutions of the type we are 
discussing here have the interpretation of being generalisations of the 
{\it supersymmetric} single-scalar solutions, in which the Page charges of 
the individual participating field strengths, whose magnitudes were 
previously required to be equal, become independent free parameters.  The 
solutions are given by \cite{lp2}
\bea
&&e^{\ft12 \epsilon \varphi_\a - d A} = 1 + \fft{\lambda_\a}{\td d} r^{-\td 
d}\ ,\nonumber\\
&&ds^2=\prod_{\a=1}^{N} \Big(1+ \fft{\lambda_\a}{\td d} 
r^{-\td d}\Big)^{-\ft{\tilde d}{(D-2)}}\, dx^\mu dx^\nu \eta_{\mu\nu} +
\prod_{\a=1}^N \Big(1+ \fft{\lambda_\a}{\td d} 
r^{-\td d}\Big)^{\ft{d}{(D-2)}}\, dy^m dy^m
\ .\label{gensol1}
\eea
We may now calculate the mass per unit $p$-brane volume and the Page 
charges for the solution, finding
\be
m = \ft14\sum_{\a=1}^N \lambda_\a\ ,\qquad P_\a = \ft14 \lambda_\a
\ .\label{mc2}
\ee
Note that in our derivation of the solutions, we assumed that the matrix 
$M_{\a\beta}$ is non-singular, which in general is the case. However, it can
be singular in two relevant cases, namely $D=5$, $N=3$ and $D=4$, $N=4$ for
the 2-form field strengths. In these cases, the analysis requires
modification; however, it turns out that (\ref{gensol1}) continues to solve
the equations of motion. 

     We have seen that multi-scalar solutions arise as generalisations of
the previous {\it supersymmetric} single-scalar $p$-brane solutions, in
which the Page charges that were previously equal become independent. 
Thus we may view the supersymmetric single-scalar solutions with $N\ge2$
participating field strengths as starting points for such
generalisations.  In the light of the findings described in section 2,
this means that multi-scalar solutions arise for 2-form field strengths,
with $2\le N\le 4$, and for 1-form field strengths with $2\le N\le 7$.
The supersymmetry of these multi-scalar solutions can determined
using  the same approach as for the single-scalar solutions, by calculating
the  eigenvalues of the \bog matrix.  A full analysis of all these
solutions and their supersymmetry is contained in
\cite{lp2}, and we shall only summarise the results here.  

For 2-form
multi-scalar solutions, the eigenvalues of the \bog matrix turn out to be
given by
\bea
N=2:&& \{p_1, u_{12}\} = \ft14 \{\lambda_1, \lambda_2 \}\ ,
\qquad {\rm for}\,\, D\le 9\ ,\nonumber\\
\mu&=&\ft12 \{ 0, \lambda_1, \lambda_2, \lambda_1 + \lambda_2 \}
\ ,\nonumber\\
N=3:&& \{u_{12}, u_{34}, u_{56} \} = \ft14 \{ \lambda_1, \lambda_2,
\lambda_3 \}\ ,\qquad {\rm for}\,\, D\le 5\ ,\nonumber\\
\mu &=& \ft12 \{ 0, \lambda_1, \lambda_2, \lambda_3, \lambda_1 + \lambda_2,
\lambda_1 + \lambda_3, \lambda_2 + \lambda_3, \lambda_1 + \lambda_2 + 
\lambda_3 \}\ ,\label{eigenvalues2f}\\
N=4:&& \{u_{12}, u_{34}, u_{56}, p_7^*\} = \ft14 \{\lambda_1,\lambda_2,
\lambda_3, \lambda_4\}\ ,\qquad {\rm for}\,\, D= 4\ ,\nonumber\\
\mu&=&\ft12 \{ 0, \lambda_1+\lambda_4, \lambda_2 + \lambda_4, \lambda_3 + 
\lambda_4, \lambda_1 + \lambda_2,
\lambda_1 + \lambda_3, \lambda_2 + \lambda_3, \lambda_1 + \lambda_2 + 
\lambda_3 +\lambda_4\}\ .\nonumber
\eea
Here a $^*$ on a Page charge indicates that the associated field strength is
dualised. Thus $p_7^*$ is the electric charge of the dualised field strength
$*{\cal F}_{\sst{MN}}^{(7)}$, and so it is the magnetic charge in terms of
the original undualised field strength ${\cal F}_{\sst{MN}}^{(7)}$. The
degeneracies of each eigenvalue in each set are equal.

     For 1-form field strengths, there exist multi-scalar generalisations
for all the supersymmetric solutions described in section 2 with $2\le
N\le 7$.  Rather than present them all here, we shall just give the
results for the case $N=7$ here, from which all the lower-$N$ cases can in
fact be derived by successively setting Page charges to zero.
The full details can be found in \cite{lp2}.  Thus we have
\bea
&&\{q_{12}, q_{34}, q_{56}, v_{127}, v_{347}, v_{567}, v^*_7\}
=\ft14 \{\lambda_1, \lambda_2, \lambda_3, \lambda_4, \lambda_4, \lambda_5, 
\lambda_6, \lambda_7 \}\ ,\nonumber\\
\mu &=&  
\{0, 
\lambda_{126}, \lambda_{135}, \lambda_{234}, 
\lambda_{147}, 
\lambda_{257}, \lambda_{367}, \lambda_{456}, \lambda_{1245}, 
\lambda_{1346}, \label{eigenvalues1f}\\
&&\ \lambda_{2356}, \lambda_{1567}, \lambda_{2467},
\lambda_{3457}, 
\lambda_{1237},
\lambda_{1234567} \}\ ,\nonumber
\eea
where we define $\lambda_{\a\beta\cdots\gamma}= \lambda_\a + \lambda_\beta +
\cdots +\lambda_\gamma$.  Note that we 
presented only one representative set of Page charges among many 
possibilities, since they all give identical eigenvalues. The degeneracy of
each eigenvalue is the same, with the total
number of eigenvalues being 32.

     For generic values of the Page charges, we see that the numbers of
zero eigenvalues of the \bog matrices for all the multi-scalar solutions
given above are the same as for their single-scalar limits in
section 2.  At certain special values of the Page charges, however, some
additional zero eigenvalues can arise, corresponding to an enhancement of
the supersymmetry.  For example when $N=2$ for 2-form solutions, we see
from (\ref{eigenvalues2f}) that by setting $\lambda_2=-\lambda_1$, the
generic 8 zero eigenvalues are enlarged to 16, implying that $\ft12$ the
supersymmetry is now preserved instead of just $\ft14$.  Since the final
eigenvalue in the set, $\ft12(\lambda_1+\lambda_2)$, is equal to twice the
mass, it follows that the solution becomes massless in this special case.
In $D=4$, this corresponds to a massless black hole.  The solution appears
to suffer from two pathologies, however.  Firstly, since one of
$\lambda_1$ or $\lambda_2$ must be negative in this special case, we can
see from (\ref{gensol1}) that there is a naked singularity.  More
seriously, perhaps, we see from (\ref{eigenvalues2f}) that some of the
eigenvalues of the \bog matrix must also now be negative.  The 
non-negativity of the \bog matrix can be proved for classical solutions 
(subject to some conditions that are evidently violated in this massless 
example), and is required at the quantum level for all acceptable states, 
since it arises as the square of the Hermitean supercharge operators.  Thus 
classical solutions with negative eigenvalues cannot form part of the true 
quantum spectrum.

     There are other examples of supersymmetry enhancements that can occur 
while maintaining the non-negativity of the \bog matrix.  For example, the 
4-scalar solution for 2-forms can lead to three inequivalent enhancements,
given by 
\bea 
\lambda_1=-\lambda\ ,&&\lambda_2 =\lambda\ ,\qquad
\mu = \ft12\{ 0_8, (\lambda_3\pm\lambda)_4, (\lambda_4\pm \lambda)_4, 
(\lambda_3 + \lambda_4)_8\}\ ,\nonumber\\
\lambda_1=-\lambda\ ,&& \lambda_2=\lambda_3=\lambda\ ,\qquad
\mu = \ft12\{0_{12}, (2\lambda)_4, (\lambda_4- \lambda)_4, 
(\lambda_4 + \lambda)_{12}\}\ ,\label{enhance2f}\\
\lambda_1=-\lambda\ ,&& \lambda_2=\lambda_3=\lambda_4=\lambda\ ,\qquad
\mu =\lambda \{0_{16}, 1_{16}\}\ ,\nonumber
\eea
where as usual the subscripts denote the degeneracies of each eigenvalue. 
Thus these three cases preserve $\ft14$, $\ft38$ and $\ft12$ of the
supersymmetry  respectively, in contrast to $\ft18$ for generic values of
the charge parameters.  Note that in these cases, although the Bogomol'nyi
matrices have  no negative eigenvalues when the supersymmetry enhancements
occur, the  metrics of the solutions still seem to have naked singularities
since one of the Page charges $\lambda_\a$ is negative. If we relax the
condition that  the eigenvalues of the Bogomol'nyi matrix should be
non-negative, then  further enhancements are possible, in which $\ft58$ or
$\ft34$ of the  supersymmetry is preserved.  (This does not
violate the classification of supermultiplets given in \cite{fs}, since
non-negativity of the commutator of supercharges was assumed there.)  
Similar supersymmetry enhancements can occur for the 1-form solutions, as 
can easily be derived from (\ref{eigenvalues1f}).

     Finally, we remark that the non-supersymmetric $p$-brane solutions that 
are obtained by reversing the signs of certain Page charges in the 
$\Delta=1', \ft45, \ft23$ and $\ft47$ single-scalar solutions also admit 
multi-scalar generalisations.  These can be summarised by presenting the 
eigenvalues of the \bog matrix for $N=4$ scalars for 2-form solutions, and 
$N=7$ scalars for 1-form solutions:
\bea
N=4:&& \mu=\ft12 \{\lambda_1,
\lambda_2, \lambda_3, \lambda_4, \lambda_{123},
\lambda_{124}, \lambda_{134},\lambda_{234}\}\ ,\nonumber\\
N=7:&&\mu = \ft12\{\lambda_{7}, \lambda_{14}, \lambda_{25}, \lambda_{36},
\lambda_{123}, \lambda_{156}, \lambda_{246}, \lambda_{345},
\lambda_{1267}, \lambda_{1357}, \lambda_{2347}, \lambda_{4567},\nonumber\\
&&\,\qquad\quad\lambda_{12457}, \lambda_{13467}, \lambda_{23567},
\lambda_{123456}\}\ .
\eea
 Note that in these cases none of
the eigenvalues is proportional to the mass, for generic $\lambda_\a$.  A 
detailed discussion of the various cases that arise by choosing special 
values for the Page charges is given in \cite{lp2}.

     To conclude, we present a table summarising the various multi-scalar
$p$-brane solutions that we have been discussing.

\bigskip

\centerline{
\begin{tabular}{|c||c|l||c|c|}\hline
{\phantom{D}Dim.\phantom{D}}&\multicolumn{2}{c||} 
{\phantom{DDDDD} 2-Forms \phantom{DDDDD}} & 
\multicolumn{2}{c|}{\phantom{DDDDD}1-Forms\phantom{DDDDD}} \\ \hline\hline
$D=10$& $\phantom{D}N=1\phantom{D}$ &\phantom{DD}$p=0,6$ & &  \\ \hline
$D=9$ & $N=2$ &\phantom{DD}$p=0,5$ & $N=1$ & $p=6$ \\ \hline
$D=8$ &       &\phantom{DD}$p=0,4$ &$N=2$&$p=5$ \\ \hline
$D=7$ &       &\phantom{DD}$p=0,3$ &     &$p=4$ \\ \hline
$D=6$ &       &\phantom{DD}$p=0,2$ &$N=3,4'$&$p=3$ \\ \hline
$D=5$ & $N=3$ &\phantom{DD}$p=0,1$ &        &$p=2$ \\ \hline
$D=4$ & $N=4$ &\phantom{DD}$p=0$   &$N=4,5,6,7$&$p=1$ \\ \hline
\end{tabular}}
\bigskip

\centerline{Table 2: Multi-scalar $p$-brane solutions}
\bigskip

\noindent  Here we list the highest dimensions where $p$-brane solutions
with the indicated numbers $N$ of field strengths first occur.  They then
occur also at all lower dimensions.

\end{document}